\documentclass[aps,prl,twocolumn]{revtex4-1}
\pdfoutput=1
\usepackage{graphicx}

\usepackage{times}
\usepackage{ulem}
\newcommand{\be}{\begin{equation}}
\newcommand{\ee}{\end{equation}}

\begin{document}

\begin{abstract}
We developed a method of precise isotope labeling to visualize the continuous growth of graphene by chemical vapor deposition (CVD). This method allows us to see in real time the growth of graphene monocrystals at a resolution of a few seconds. This technique is used to extract the anisotropic growth rates, the formation of dendrites, and the dependence on adsorption area of  methane on copper. We obtain a physical picture of the growth dynamics of graphene and its dependence on various parameters. Finally, our method is relevant to other CVD grown materials. 
\end{abstract}

\title{Time Evolution of the Growth of Single Graphene Crystals and High Resolution Isotope Labeling}
\author{Eric Whiteway$^{\ast}$, Wayne Yang, Victor Yu, Michael Hilke}
\affiliation{Department of Physics, McGill University, Montr\'eal, Canada H3A 2T8}

\date{\today}
\maketitle

\section{Introduction}
Visualizing the growth of materials has been of interest for a long time. This can be as simple as watching water crystals grow on a windshield or more complex such as observing the growth of proteins in a liquid environment. In most cases this is done in-situ via optical means. However, at the atomic level this becomes more difficult and other techniques need to be used such as LEED or scanning probe techniques \cite{oda11}. Typically, crystals grow in an ``unfriendly" environment, with very high temperatures or extreme pressures, which doesn't permit an in-situ observation. In our work we developed a method to continuously image and recreate the crystal growth ex-situ, i.e., after the crystal is fully grown. We have applied this technique to graphene crystals, which is a fascinating system by itself, but the method is generalizable to many other materials. This is achieved by varying the relative ratio of the isotopes continuously during the growth process. The isotope concentration will therefore provide a unique label determining when the material was grown. This method is similar to carbon dating, where the isotope concentration of radio active carbon decays over time and therefore allows the dating of organic materials \cite{arnold49age}. The difference here, are the time scales involved (seconds) versus hundreds of years and the spatial dependence, which allows us to correlate the position with time.

The ability to visualize the growth of atomic materials has important implications, particularly with the recent development and potential of hybrid materials beyond graphene \cite{novo04,butler13progress}, where the atomic layering and growth history is an important factor. For graphene, in particular, this method not only gives new insights on the growth dynamics, like the formation of dendrites, of nucleation sites or additional layers as discussed here, but the varying isotope method can also be used to label certain areas of the crystal without changing its electronic properties. Other applications include the possibility to tune the nuclear spin density \cite{Wojtaszek14absence}, the ability to do phonon engineering \cite{balandin2014phonon}, reduce thermal conductivity due to phonon disorder \cite{chen2012thermal}, determine chemical and vibrational properties of different layers, \cite{li13}, increase sensitivity to determine specific phonon modes \cite{bern12}, and others \cite{kalbac2012control,wang14raman,frank14the}.

Here we provide new results organized along three main themes: {\bf continuous labeling} and determination of the sensitivity of the method, which can provide close to hundred independent labeling or time steps; {\bf dynamic growth} or the making of movies and time snapshots of the growth, in particular the fractal growth of graphene; {\bf growth rates} and the dependence of the growth rates on anisotropy, dendricity and growth diffusion area.

Graphene is a two dimensional material which has attracted a great deal of attention for its unique physical and electronic properties\cite{novo05,geim09,coop12}. Originally isolated by mechanical exfoliation of graphite\cite{novo04}, more scalable methods necessary for production of large scale graphene layers and industrial applications include chemical vapour deposition\cite{yu08,li09} and graphitization of SiC\cite{emts09}.
Chemical vapour deposition (CVD) of graphene sheets on commercial polycrystalline copper foils is a leading candidate for the production of large high quality graphene films. The main challenge in producing high quality graphene by CVD is controlling the nucleation density of graphene crystals which in turn limits the size of individual crystals. Grain boundaries where different misoriented graphene crystals come together are known to reduce the mechanical strength and electronic properties of graphene\cite{yu11,huan11}. As a result it is crucial to be able to produce large monocrystal graphene sheets. Since the pioneering work of Li et al. in graphene CVD on copper foils \cite{li09} a great deal of work has been done in determining the optimal conditions to produce large single crystal graphene and understanding the role of various growth parameters in the CVD process\cite{li10}. This has led to large gains in the size and quality of single crystal graphene flakes which can be produced\cite{chen13,hao13} up to mm and cm scales. Analyzing and understanding the growth dynamics of CVD graphene, is one of the main contributions of this work, where we start by describing a new labeling method.
\section{High resolution continuous labeling method}
\begin{figure*}[ptb]

\includegraphics[width=5.8in]{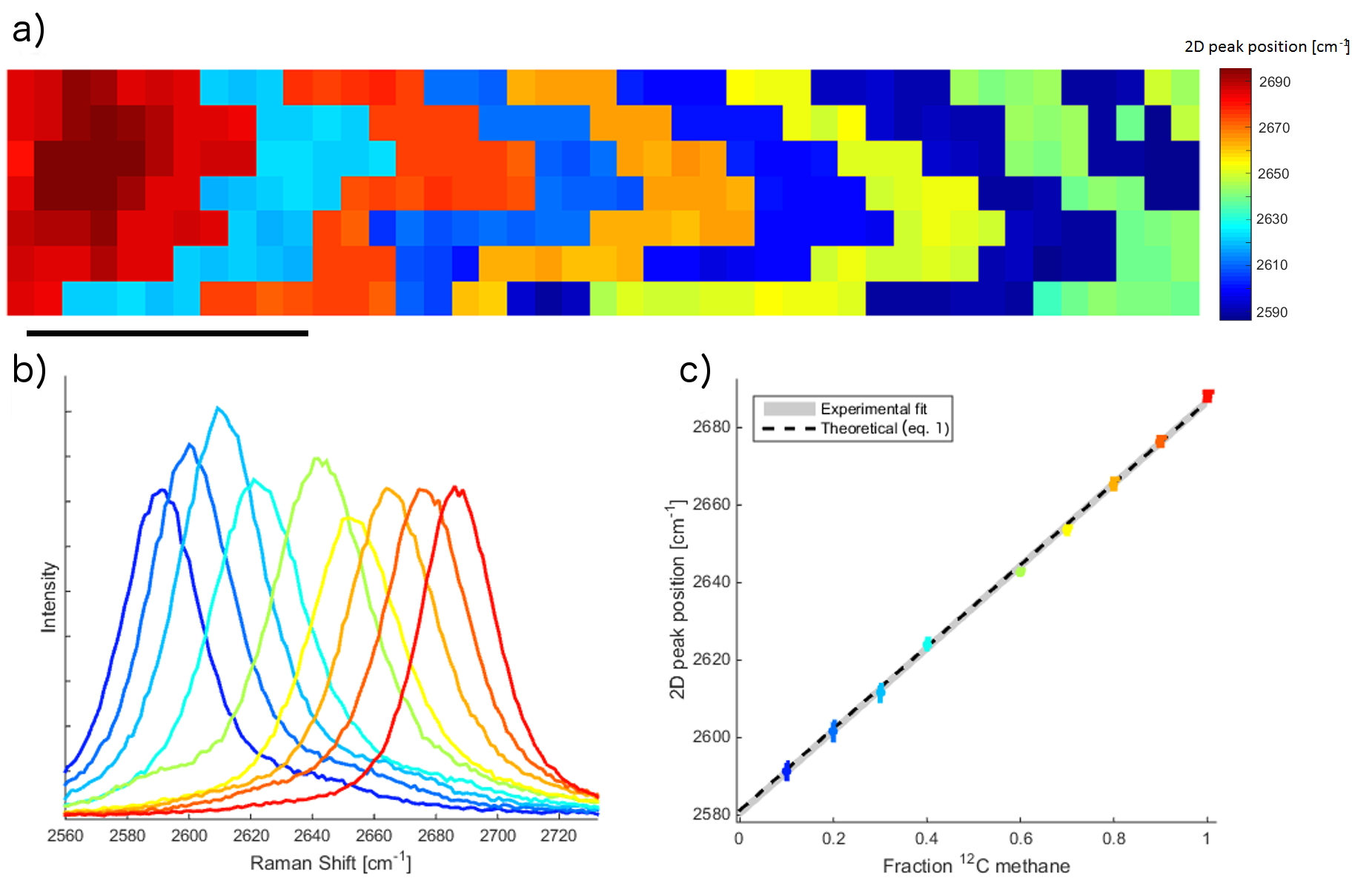}
\caption{a) Raman map showing 2D position for the single sample containing different ratios of $^{12}$C:$^{13}$C graphene. Scale bar is 10 $\mu$m 
b) Representative Raman spectra for the different isotopic ratios showing the 2D peak. c) We observe a linear shift in position dependent on the amount of $^{13}$C vs. $^{12}$C present during growth. The position of the 2D peaks shown as a function of $^{12}$C fraction. Solid grey line indicates a linear curve fit to the experimental data, with the line thickness representing a 95\% confidence interval. The dashed line indicates the theoretical curve given by equation 1.}
\label{ratios}
\end{figure*}

In general, the possibility to label a material locally is of importance in a number of applications. Indeed, it allows to encode information, without modifying the basic properties. In the case of graphene, this can be particularly useful in areas where transparency, organic composition or low mass is important, such in labeling electronic devices or biomolecules. The labeling is done by using different relative concentrations of isotopes, which can then be used for encoding or transparent labeling. The detection is achieved by Raman spectroscopy, which is used to characterize graphene \cite{mala09} and other graphitic materials\cite{jori11}. Here we use a combination of precisely controlled isotopic methane flow ($^{12}$C and $^{13}$C) and Raman mapping to probe the growth dynamics of graphene crystals in much greater detail than is possible when using only a binary on/off isotopic variation method \cite{li092}.

Several samples were prepared using pure $^{12}$CH$_4$, pure $^{13}$CH$_4$ and various mixes. In the $^{13}$C carbon isotope graphene we observe a downshift in Raman peak position by a factor of approximately $\sqrt{12/13}$ as a result of the change in the mass of the carbon atoms\cite{li09,bern12}.
A sample was prepared where the ratio of $^{12}$C:$^{13}$C methane was modified stepwise throughout the growth in 10\% intervals to give ratios of 9:1, 8:2, etc. A Raman map of this sample is shown in figure \ref{ratios}c where we observe bands of graphene with different 2D peak positions. Plotting the position of the 2D peak for each of these bands we find and excellent agreement with the predicted linear behaviour where the position of the peak is a function of the average mass of carbon atoms in a given region.
Taking x to be the fraction of $^{12}C$ and $\Delta\omega= \omega_{2D}(^{12}C)-\omega_{2D}(^{13}C)$ we have
\begin{equation}\omega_{2D}(t)=\omega_{2D}(^{13}C)+x\Delta\omega
\label{eq1}
\end{equation}
The fit is done to a linear function with $\omega_{2D}(^{12}C)$ and $\omega_{2D}(^{13}C)$ as free parameters and gives values of 2686.7 cm$^{-1}$ and 2580.6 cm$^{-1}$ respectively.

The theoretical line plotted in figure \ref{ratios}c shows the expected result taking $\omega_{2D}(^{12}C)$ as 2687 cm$^{-1}$. The excellent agreement quantifies the precision with which the isotopic ratio of our graphene during CVD growth and extract this value by Raman spectroscopy of our samples. The average standard deviation for the peak position for a given concentration is $\sigma=1.4$ cm$^{-1}$. This translates into a labeling resolution of
\begin{equation}
R_l=\sigma/\Delta\omega\simeq 1.3*10^{-2}
\end{equation}
The high resolution allows the labeling of $1/R_l\simeq 80$ regions independently. Moreover, most physical properties are not affected by the labeling, except for Raman process. This can be used for a variety of applications, such as encryption, variable nuclear spin concentration, or even isotope concentration detection. In what follows, we will use it in order to gain insight into the growth mechanism of graphene.

\section{Dynamic Imaging}
\begin{figure*}[pt]

\includegraphics[width=5.8in]{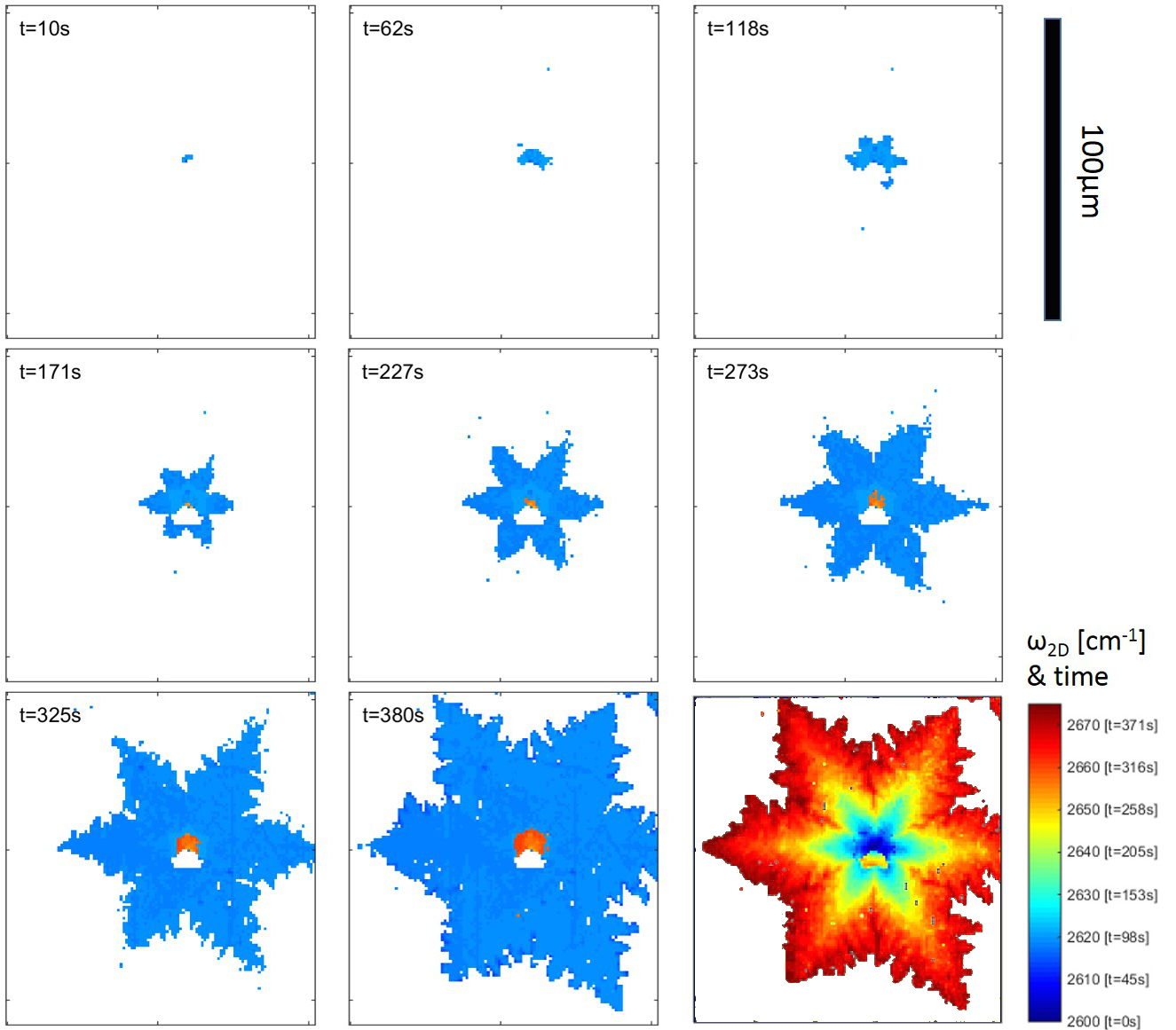}
\caption{Visualizing the growth of a graphene crystal: Snapshots of the graphene growth showing the combined intensity of the 2D peak as a function of time. Intensities are scaled to differentiate between layers 1 and 2. Last square: Raman maps of a dendritic graphene crystal indicating the Raman shift of the 2D peak and associated time from the peak position to time correspondence. Only the layer 1 peak position is shown.}
\label{gcontime}
\end{figure*}
In situ observation of the CVD growth process is technically demanding\cite{oda11} and as a result the growth mechanisms of the chemical vapour deposition process are not fully understood. Most of the understanding of the CVD growth process is the result of ex situ analysis of samples prepared by CVD across a range of conditions\cite{li10}. The use of Isotopic methane is a useful tool that has been used to characterize the growth of polycrystalline sheets \cite{li09,li092} and graphene single crystals. Typically this has been done by using alternating flow of $^{12}$C and $^{13}$C methane to produce graphene which displays a tree-ring like growth pattern \cite{li09}. Here, several graphlocon samples were prepared with the ratio of $^{12}$C to $^{13}$C varying linearly as a function of time. The sample shown in figure \ref{gcontime} was first annealed in hydrogen at 1073$^\circ$C for 4 hours to prepare the surface and then graphene was deposited during a 400 second growth phase with a combined methane flow of 1.2 sccm. Since the Raman peak position is proportional to the isotopic concentration (figure \ref{ratios}), we can then extract the expected 2D peak position at any given time using eq. 1.

The growth conditions employed for this sample were designed to give an incomplete growth with dendritic graphlocons \cite{mass13}. A Raman map was taken with grid spacing of 900 nm. A spectrum was taken at each point on the grid with a collection time of 1s. The position,width and intensity of the 2D peak were then extracted. We observe that the 2D position is lowest at the center of the graphlocon and increases as we move towards the edges. This is consistent with a growth that begins with primarily $^{13}$C methane, with the $^{12}$C concentration increasing with time. The analysis could equally be performed using the G or other Raman graphene peak, however the 2D peak was chosen for the combination of large amplitude and high Raman shift, which enable the position of the peak to be extracted with greater precision and lower collection time relative to other peaks.
We also observe in the center of the graphene crystal a region of bilayer graphene, characterized by a double Raman peak structure. The growth times for each graphene layer are calculated from the position of the two independent 2D peaks observed. In fact, we observed two types of bilayer growth, either Bernal stacked or not, sometimes even in the same crystal. In the case of Bernal stacking, the 2D Raman peaks of each layer hybridize \cite{ferr06} to form one peak corresponding to the average mass between the two layers, while in the non-Bernal stacking, the two Raman peaks identify the frequencies corresponding to the respective atomic mass of each layer, with a weak enhancement due to graphene enhanced Raman scattering \cite{ling2010first}.  This is indicated in figure \ref{gcontime} where the bilayer is shown in orange. In the last square of figure \ref{gcontime} we show only the contribution of the first Raman peak corresponding to the first layer. 

This linear shift in peak position allows us to extract an absolute time scale for the progression of the crystal formation. We can visualize the growth of the graphlocon as a function of time by considering the points in the Raman map with $\omega_{2D}< \omega_{2D}(t)$. In figure \ref{gcontime} we show snapshots of the graphlocon at 8 different points in time. In the supplementary material the full movie of the graphene growth is shown.

\section{Growth rates: Anisotropy, dendricity, and diffusion area}
\begin{figure*}[ptb]
\includegraphics[height=1.6in]{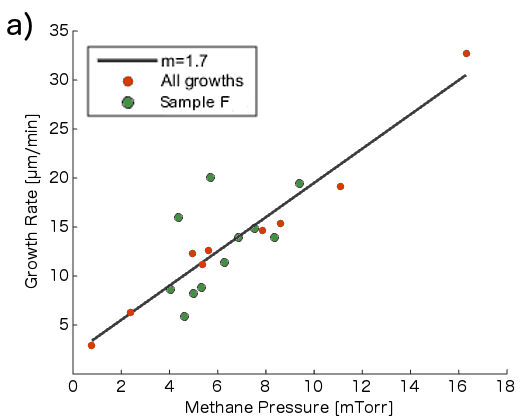}
\includegraphics[height=1.6in]{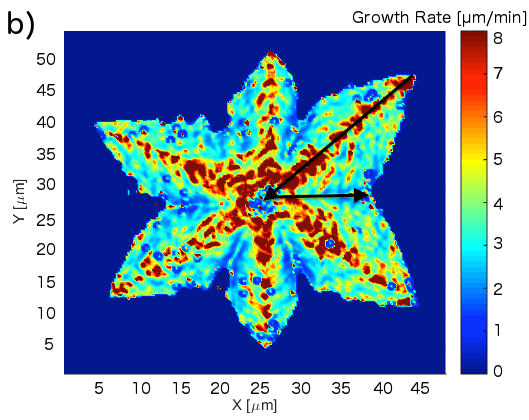}
\includegraphics[height=1.6in]{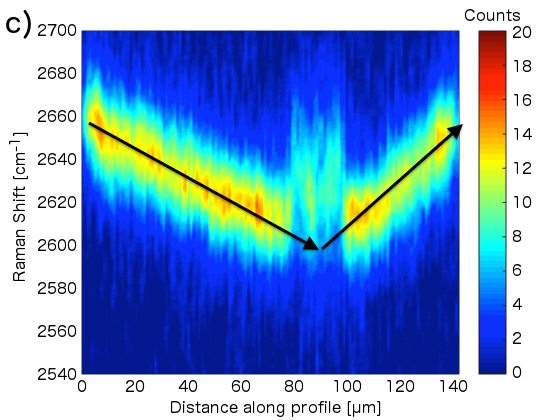}
\caption{a) Effect of methane pressure on growth speed. b) Magnitude of the growth velocity shows an increased rate of growth along the six primary arms of the graphene crystal. c) Colour map indicating the spectral intensity as a function of Raman shift and position, taken along the fast and slow profiles of the crystal as indicated in b.}
\label{angdep}
\end{figure*}
It’s clear from the isotope labelling that the graphene crystal grows from the center outwards. This is consistent with previous studies on CVD graphene growth \cite{li092,li11,vlas11}. In some of these studies the crystal symmetry was also determined and various shapes have been observed, including four-fold and six-fold symmetries, depending on the growth conditions and substrate \cite{wu13}. However, the local growth rates have not yet been analyzed. Using our data from the method described in section 3 we are able to extract the growth rates and directions at every point in the crystal.  

\subsection{Anisotropy}

The growth anisotropy is not affected by the growth stage and is constant over time during the growth and independent of the size of the crystal. The radial growth is indeed linear with radius (see figure 3c) and therefore time, since we chose $x\sim t$. Great care was taken to calibrate the $^{12}C$ and $^{13}C$ methane flows, so that the total methane flow is constant over time to allow for a precise determination of the time dependence. To analyze the anisotropy in more detail, we compare the growth rate in different directions for a weakly dendritic graphene crystal, shown in figure \ref{angdep}b. This sample was prepared with a combined $CH_4$ flow rate of 1.2 sccm and a $H_2$ flow rate of 80 sccm, during a 3 minute growth phase. The highest growth rates are clustered along 6 arms spaced by roughly 60$^{\circ}$. By taking profiles along the fastest and slowest directions we observe that the average growth rate is roughly doubled along the fast profiles when compared to the slow ones, 9.0 $\mu$m/min vs 4.7 $\mu$m/min.

The growth rate was extracted for growths performed across a range of methane gas flows. Comparing the calculated partial pressure of methane gas with the growth rate we observe a clear linear relationship. Figure \ref{angdep}a shows the relationship. In all cases this data was measured along the fastest growth direction, and includes samples prepared using the $^{12}$C/$^{13}$C technique discussed in this article as well as pure $^{12}$C samples where the growth rate is calculated from the diameter of the graphlocon and the growth time.
To probe whether this relationship holds within the same single crystal, we synthesized sample F using a variable overall methane flow as well as shifting the $^{12}$C/$^{13}$C ratio and the same range of growth rates as those extracted from individual growths was obtained.

\subsection{Fractal shape and dendricity}
\begin{figure*}[ptb]
\includegraphics[width=3.5in]{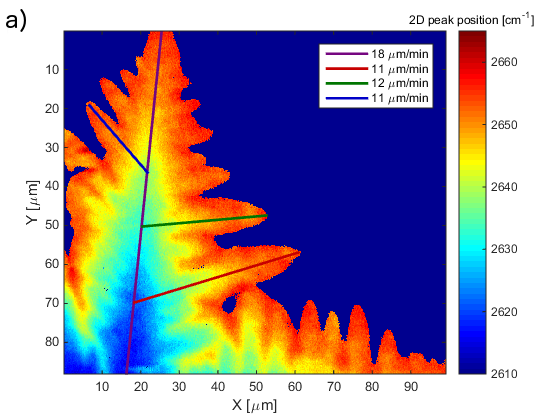}
\includegraphics[width=3.5in]{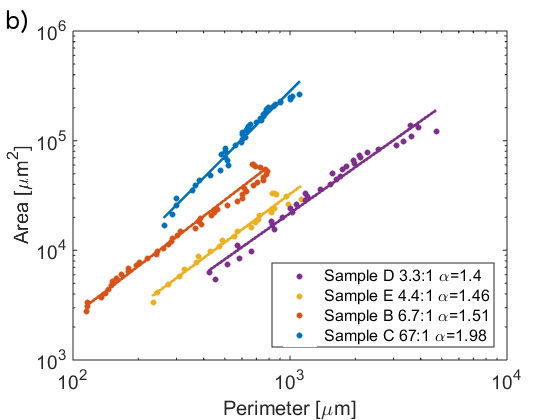}
\caption{a) Extracting the average growth rate of the primary arm and several dendrites of a large graphlocon b) Perimeter and area of the graphlocon as a function of time for several samples prepared with different growth conditions. Data points are offset on the y-axis to show the difference in slope, such that the area of samples D, E and B are scaled by factors of 8,4 and 2 respectively.}
\label{dendritesmap}
\end{figure*}
In ref. \cite{mass13} it was shown that fractal graphene crystals can be synthesized using CVD. Fractal shapes can lead to an enhanced broadband efficiency in antenna designs\cite{rumsey1966frequency} and the corresponding increased edge to surface ratio,  can lead to an enhanced reactivity for chemical processes at the edge \cite{jeon2012large}. Pushing the anisotropy further could even lead to the synthesis of quasi one dimensional arms or graphene nano-ribbons. 

A large dendritic sample was synthesized using low pressure and low gas flow rates. The copper foil was sonicated in acetic acid for 30 minutes prior to growth in order to suppress graphene nucleation and allow larger domains to form. A high resolution Raman map shows details of the formation of the dendritic arms of the graphene crystal. The growth rates were  extracted along the primary arm of the crystal as well as several dendrites and shows the dendrites forming concurrently with the main arm. The average linear growth velocity is extracted by fitting the distance along each profile to the time corresponding to the 2D peak position at each point on the profile. We see that the main arm has a velocity of ~18 $\mu$m/min while the dendrites on the side arms have a slower average velocity 11-12 $\mu$m/min even though this is along the “fast” direction. If we slice our data more finely we can see that many of the dendrites in fact show an accelerating growth rate. This increasing growth rate as the dendrites become more pronounced suggests that the growth is dependent on the local available copper surface, which we discuss in more detail in the next section. A similar analysis was performed for several different samples. All samples were produced using a similar low pressure CVD growth, but with small variations on the conditions. From the Raman maps of these samples we can extract the area and perimeter of the graphlocon as a function of time.

We consider the scaling exponent $\alpha$ defined by $A \sim P^\alpha$, where $A$ is the area and $P$ the perimeter. A compact hexagon would give $\alpha=2$, whereas for a more dendritic shape the value should be lower \cite{scott06}. In previous works, various scaling exponents have been extracted, ranging between 1.43 and 2 for different flakes under different growth conditions and particularly sensitive to growth temperature \cite{mass13}. In these works, $\alpha$ could only be extracted once at the end of the growth for every grown crystal. Here, on the other hand, we are able to determine how $\alpha$ evolves during the growth process. Remarkably, we see that the scaling exponent doesn’t change as a function of time as shown below. This means that the growth conditions determines $\alpha$, which is time-invariant, while the growth time simply determines the final size of the crystal. Hence, for a dendritic crystal, the growth rate in the fastest direction is not constant, but $\alpha$ is.

In figure \ref{dendritesmap}b we show the area versus perimeter of 4 different graphene crystals. A powerlaw fit gives values of $\alpha$ ranging from 1.4 to 1.98 for the scaling exponent. The range of validity of a single powerlaw extends over the entire size of the crystal, demonstrating the time invariance of $\alpha$ during the growth process. We also note that the fractal dimension increases with an increasing ratio of hydrogen to methane during the growth. In one case (sample D) we obtain a value of $\alpha$ considerably larger than for the other samples. This sample, also shown in figure \ref{angdep}b, was prepared with a much larger hydrogen flow of 80 sccm when compared to all the other samples, which ranged from 4 to 12 sccm $H_2$ and leads to fewer dendrites and $\alpha\simeq 2$.  

\subsection{Diffusion growth area}
Earlier studies have already examined the time dependence of dendrite growth and merging in graphene and found that growth is faster at the tip of the dendrite and slower for regions in between dendrites and slower at the end of the growth\cite{li11}. However these studies lack the time resolution to conclusively determine the factors affecting the growth rate. To shed light on this process, we show that the growth rate correlates with the locally available copper surface and argue that the dendritic growth is limited by the diffusion of the adsorbed C species on the available copper surface. We estimate the related diffusion rate to be 6 $\mu$m as discussed below.
\begin{figure*}[t]
\includegraphics[width=1.55in]{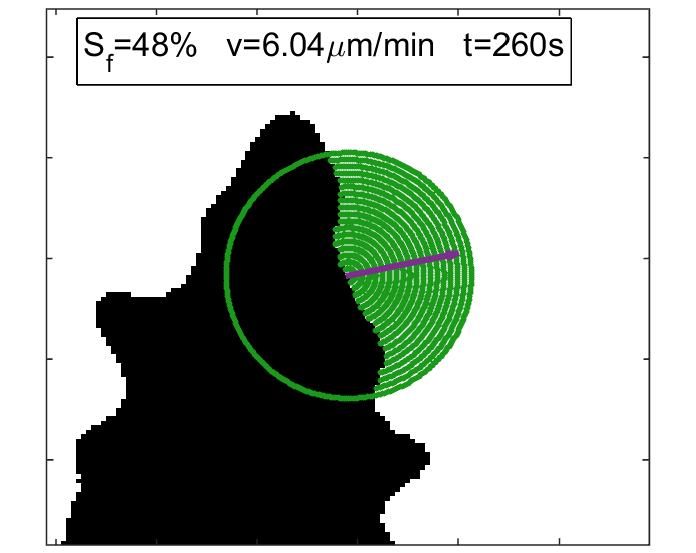}
\includegraphics[width=1.55in]{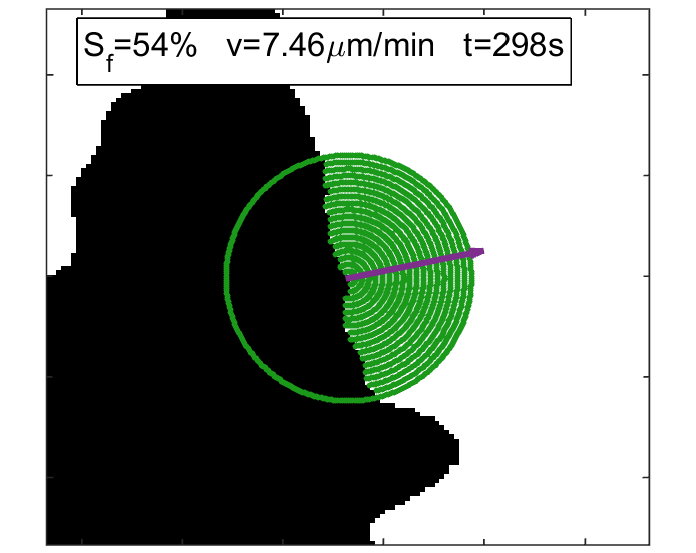}
\includegraphics[width=1.55in]{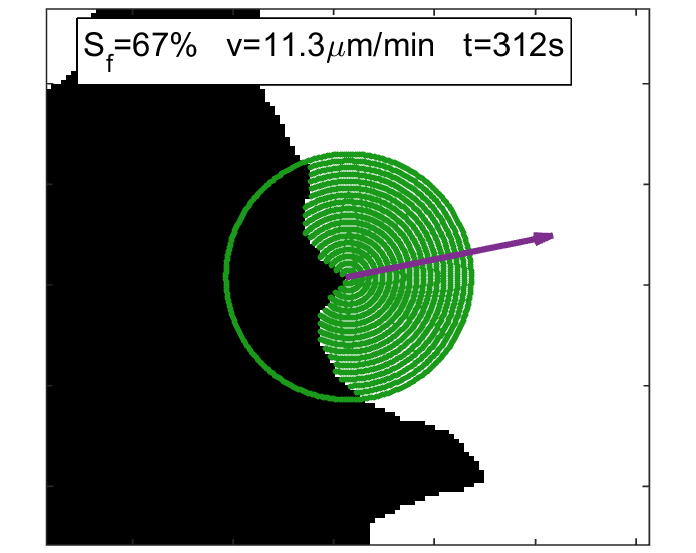}
\includegraphics[width=1.55in]{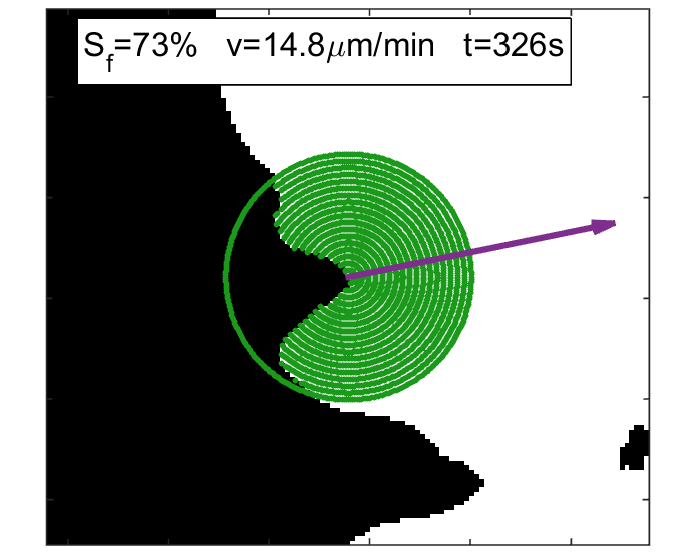}
\includegraphics[width=6.2in]{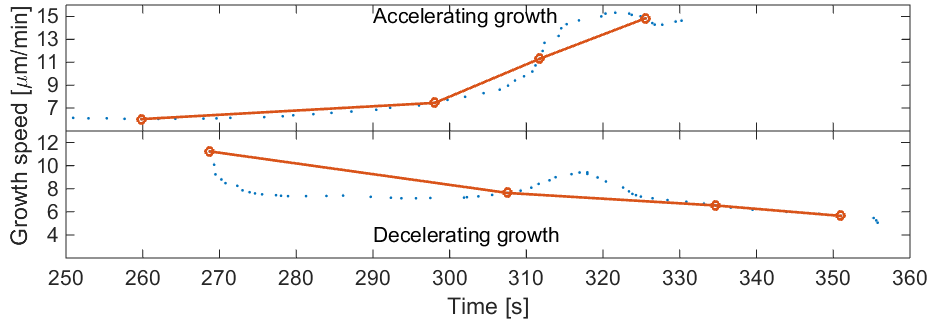}
\includegraphics[width=1.55in]{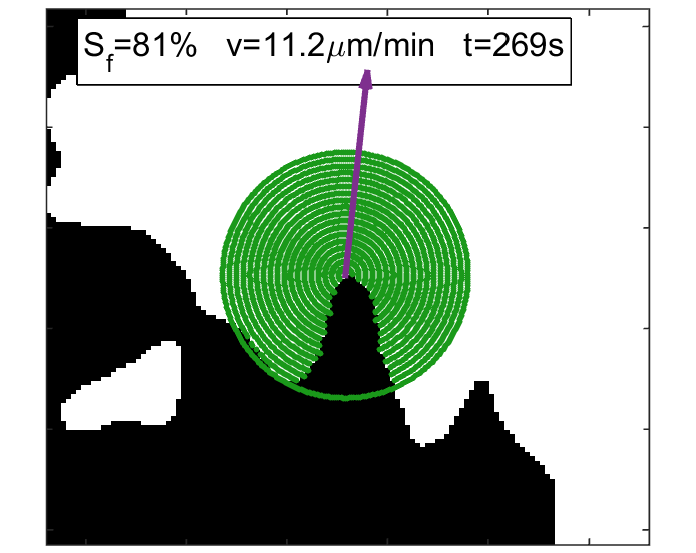}
\includegraphics[width=1.55in]{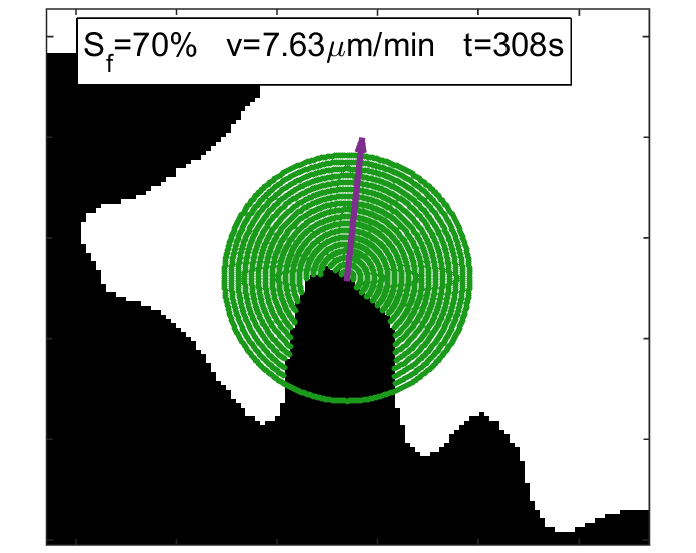}
\includegraphics[width=1.55in]{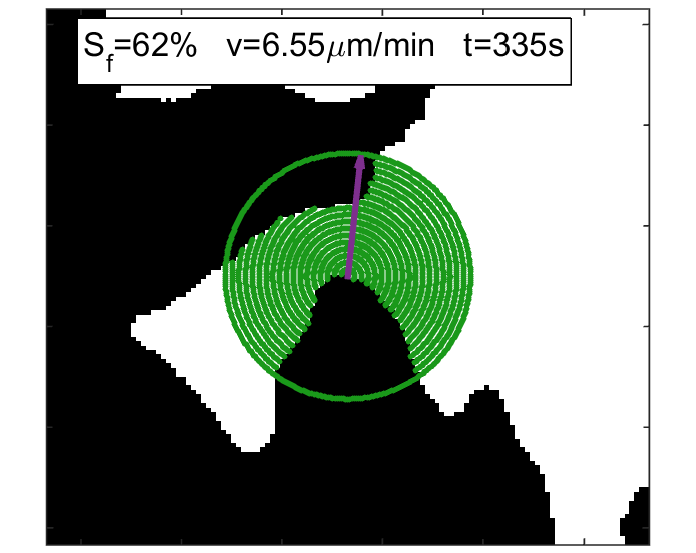}
\includegraphics[width=1.55in]{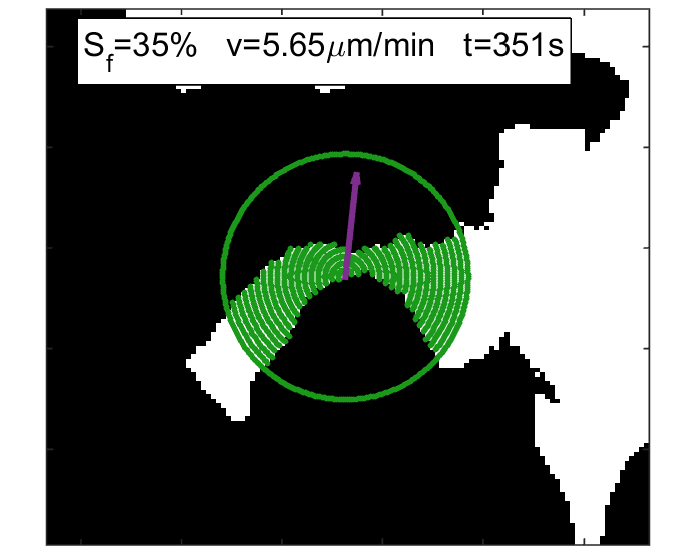}

\caption{
Top: Example of an accelerating dendrite. Filled green circle indicates the free copper surface , $S_f$, while the arrow indicates the magnitude of the growth velocity, v.  Middle: Growth velocity as a function of time, with large circles indicating the snapshots shown. Bottom: an example of a decelerating dendrite. A full movie can be seen in the supplementary material }
\label{dva}
\end{figure*}

To gain insight on the locally available copper surface and the growth rate we extracted the growth rate along many different profiles along the edge by focusing on the fastest growth direction. From the high resolution raman map we can obtain for any given time $t$, the exact shape of the graphlocon as well as the growth rate in any direction. In figure \ref{dva} we see two examples illustrating results for different dendrites. The top line shows the behaviour of a dendrite growing into free space. As the dendrite grows the shape becomes more pronounced and the available copper surface increases. This results in an increasing growth rate. The second line shows a dendrite growing into occupied space. In this case as the dendrite grows the available copper surface is decreasing and the growth rate is limited by the lack of available surface. This results in a decreasing growth rate. The correlation between the locally available copper surface and the growth rate was averaged across 20 dendrites. We observe an approximately linear relationship above 30\% available surface and a roughly constant growth rate of 5$\mu$m/min below 30\%.

In order to determine the approximate radius of the diffusion area we first assume that the growth rate is a linear function of free surface area over the range of 30\%-70\%. We fit the data in that range considering radii from 1 to 20 $\mu$m and extract the coefficient of determination $r^2$ for the linear regression. The radius corresponding to the maximum $r^2$ determines our diffusion radius, which we find to be approximately 6 $\mu$m. This value is similar to the observed spacing of dendrites seen in figure \ref{dendritesmap}a. The mechanism emerging from this, is that the methane is adsorbed on the copper surface and then diffuses on that surface. Once a graphene nucleation has started, the edge of the crystal requires more methane while at the same time expelling the accumulated hydrogen from the copper induced catalytic conversion of methane to carbon. The rate of this reaction depends on the local concentration of hydrogen and methane gases. The more methane is available within its diffusion area, the faster the growth of graphene. 

\section{Conclusions}

To conclude, we have developed a technique, which allows to image dynamically the growth of graphene mono-crystals in addition for providing a tool to label different parts of the crystal. This imaging technique enables the assessment of the various anisotropic and non-monotonous growth rates responsible for the dendritic and fractal growth of graphene. These graphene monocrystals grow from nucleation sites, which are typically impurities or deformations in the copper substrate. 
We observed that the graphene nucleation is time dependent with individual crystals nucleating at different times independent of other factors. Smaller single layer graphene crystals form as a result of delayed nucleation, but the mechanism and growth rate are the same for both smaller and larger crystals. Finally, the dynamic imaging allows us to draw a precise picture of the CVD growth process, including the relevant diffusion area of the adsorbed species responsible for the growth. Moreover, this technique can easily be expanded to other materials whenever source elements with different isotopes are available, including common elements such as silicon and nitrogen.  

\section{Acknowledgments:}
We thank NSERC and FQRNT for financial assistance. We would also like to thank Mr. James Medvescek for his valuable insights into entrance lengths and laminar flow development. As well as Mr. Robert Gagnon for the technical assistance and Dr. Samir Elouatik for the Raman spectroscopy.

\bibliographystyle{Science}
\bibliography{growth}
\end{document}